
\magnification = 1200
\def\lapp{\hbox{$ {
\lower.40ex\hbox{$<$}
\atop \raise.20ex\hbox{$\sim$}
}
$}  }
\def\rapp{\hbox{$ {
\lower.40ex\hbox{$>$}
\atop \raise.20ex\hbox{$\sim$}
}
$}  }
\def\barre#1{{\not\mathrel #1}}
\def\krig#1{\vbox{\ialign{\hfil##\hfil\crcr
$\raise0.3pt\hbox{$\scriptstyle ?$}$\crcr\noalign
{\kern-0.02pt\nointerlineskip}
$\displaystyle{#1}$\crcr}}}
\def\upar#1{\vbox{\ialign{\hfil##\hfil\crcr
$\raise0.3pt\hbox{$\scriptstyle \leftrightarrow$}$\crcr\noalign
{\kern-0.02pt\nointerlineskip}
$\displaystyle{#1}$\crcr}}}
\def\ular#1{\vbox{\ialign{\hfil##\hfil\crcr
$\raise0.3pt\hbox{$\scriptstyle \leftarrow$}$\crcr\noalign
{\kern-0.02pt\nointerlineskip}
$\displaystyle{#1}$\crcr}}}

\def\Tr{\,{\rm Tr }\,}

\def\g5{\gamma_5}

\def\lp1{{\cal L}_{\pi N}^{(1)}}
\def\lp2{{\cal L}_{\pi N}^{(2)}}
\def\lp3{{\cal L}_{\pi N}^{(3)}}

\font\btr=cmr10 scaled\magstep2
\topskip=0.60truein
\leftskip=0.18truein
\vsize=8.8truein
\hsize=6.5truein
\tolerance 10000
\hfuzz=20pt

\baselineskip 14pt plus 1pt minus 1pt
\pageno=0
\centerline{\bf {\btr APPLIED CHIRAL PERTURBATION THEORY}}
\vskip 48pt
\centerline{{\btr Ulf-G. Mei{\ss}ner}\footnote{$^\ddagger$}
{Address after October 1st, 1994:
 Universit\"at Bonn, Institut f\"ur Theoretische Kernphysik,
Nussallee 14--16, D--53115 Bonn, Germany}}
\vskip 16pt
\centerline{{\it Centre de Recherches Nucl\'{e}aires et Universit\'{e}
Louis Pasteur de Strasbourg}}
\centerline{\it Physique Th\'{e}orique,
BP 28Cr, 67037 Strasbourg Cedex 2, France}
\vskip 1.0in
\centerline{ABSTRACT}
\medskip
\noindent I consider some selected topics in chiral perturbation theory (CHPT)
as probed at colliders such as DA$\Phi$NE.
 Emphasis is put on processes involving pions in the isospin
zero S-wave which require multi-loop calculations. These include the scalar
form factor of the pion, two--photon fusion into pion pairs and $K_{\ell
4}$--decays. The physics of the chiral anomaly is briefly touched upon
\medskip
\vskip 1.0in
\centerline{INVITED LECTURE}
\centerline{SUMMER SCHOOL ON HADRONIC ASPECTS OF COLLIDER PHYSICS}
\centerline{Zuoz, Switzerland, August 1994}
\vfill
\noindent CRN--94/53  \hfill September 1994
\vskip 12pt
\eject
\baselineskip 14pt plus 1pt minus 1pt
\noindent{\bf 1. INTRODUCTION}
\bigskip
This talk will be concerned with certain aspects of the standard model in
the long--distance regime. I will argue that there exists a rigorous
calculational scheme and that plenty of interesting and {\it fundamental}
problems await a solution. Particular emphasis is put on reactions to be
measured at the $\phi$--factory DA$\Phi$NE or at other places where intense
kaon fluxes are available (like e.g. Brookhaven). There are many accurate
predictions of chiral perturbation theory which await detailed tests. The
bottom line is that these low energy reactions will tell us about our
understanding of the mechanism of spontaneous chiral symmetry breaking in QCD
and also lead to a rich phenomenology.

Our starting point is the observation that in
the three flavor sector, the QCD Hamiltonian can be written as
$$\eqalign{ H_{\rm QCD} &= H_{\rm QCD}^0 + H_{\rm QCD}^I \cr
H_{\rm QCD}^I &= \int d^3x \lbrace m_u \bar u u + m_d \bar d d + m_s \bar s s
\rbrace
\cr} \eqno(1)$$
with $H_{\rm QCD}^0$ symmetric under chiral SU(3)$_L \times$ SU(3)$_R$.
On a typical hadronic scale, say $M_\rho = 770$ MeV, the current
quark masses $m_q = m_u, m_d, m_s$ can be considered as perturbations. The
chiral symmetry of the Hamiltonian is spontaneously broken down to its
vectorial subgroup SU(3)$_V$ with the occurence of eight (almost) massless
pseudoscalar mesons, the Goldstone bosons
($\varphi = \pi^+,\pi^0,\pi^-,K^+,K^-,K^0,\bar{K}^0,\eta$)
$$ M^2_\varphi = m_q \, B + {\cal O}(m_q^2)  \eqno(2)$$
with $B = -<0|\bar q q|0>/F_\pi^2$ and $F_\pi \simeq 93$ MeV the pion decay
constant. Clearly, from eq.(2) one gets immediately some information about
the ratios of the light quark masses, $m_u /m_d = 0.66$, $m_d / m_s = 1/20.1$
and $2m_s / (m_u + m_d) = 24.1$ (modulo higher order and electromagnetic
corrections, see also section 2). In the
confinement (long-distance) regime, the properties of the standard model
related to this symmetry can be unambigously worked out in terms of an
effective Lagrangian,
$${\cal L}_{\rm QCD} = {\cal L}_{\rm eff} [U, \partial_\mu U, \ldots, {\cal M}]
\eqno(3)$$
with ${\cal M} = {\rm diag}(m_u,m_d,m_s)$ the quark mass matrix and the
Goldstone bosons are collected in the matrix-valued field $U(x) = \exp \lbrace
i \sum_{a=1}^8 \varphi_a (x) \lambda^a / F_\pi \rbrace $.
 Of course, there is an infinity of
possibilities of representing the non-linearly realized chiral symmetry. While
the QCD Lagrangian is formulated in terms of quark and gluon fields and the
rapid rise of the strong coupling constant $a_S (Q^2)$ with decreasing $Q^2$
forbids a systematic perturbative expansion, matters are different for the
effective field theory (EFT) based on the effective Lagrangian (3). It can be
written as a string of terms with increasing dimension,
$${\cal L}_{\rm eff} = {\cal L}_{\rm eff}^{(2)} +
{\cal L}_{\rm eff}^{(4)} + {\cal L}_{\rm eff}^{(6)} + \ldots \eqno(4) $$
if one counts the quark masses as energy squared. To lowest order, the
effective Lagrangian contains two parameters, $F_\pi$ and $B$. It is worth to
stress that $B$ never appears alone but only in combination with the quark mass
matrix, alas the pseudoscalar meson masses. Consequently, any matrix--element
$<ME>$ for the interactions between the pseudoscalars can be written as
$$<ME> = c_0 ({E \over \Lambda})^2 +
\biggl[ \sum_{i=1}^n (c_{1i}) + ({\rm non-local}) \biggr]
({E \over \Lambda})^4 + {\cal O}({E \over \Lambda})^6 \eqno(5)$$
This is obviously an energy expansion or, more precisely, a simultaneous
expansion in small external momenta {\it and} quark masses.
The first term on the r.h.s. of (5) leads to  nothing but the well--known
current algebra results, the pertinent coefficient $c_0$ can be entirely
expressed in terms of $F_\pi$, the Goldstone masses and some numerical
constants. As one of the most famous examples I quote Weinberg's result for
the S-wave, isospin zero scattering length [1],
$$ a_0^0 = {7 M_\pi^2 \over 32 \pi F_\pi^2} \eqno(6)$$
which is such an interesting observable
because it vanishes in the chiral limit $M_\pi \to 0$.
At next-to-leading order, life is somewhat more complicated. As first shown by
Weinberg [2] and discussed in detail by Gasser and Leutwyler [3], one has to
account for meson loops  which are naturally generated by the interactions.
These lead to what I called "non--local" in (5).  In
fact, it can be shown straightforwardly that any N--loop contribution is
suppressed with respect to the leading order result by $(E / \Lambda)^{2N}$.
At ${\cal O}(E^4)$, the loop contributions do not introduce
any new parameters. However, one also
has to account for the contact terms of dimension  four which are
accompanied by a priori unknown coupling constants (the $c_{1i}$ in (5)). These
so--called low--energy constants  serve to renormalize the infinities
related to the pion loops.
Their finite pieces are then fixed from some experimental input. In the case of
flavor SU(2), one has $n = 7$. Two of these constants are related to
interactions between the pseudoscalars, three to quark mass insertions and the
remaining two have to be determined from current matrix elements.
The inclusion of gauge boson couplings to the Goldstone
bosons is most simply and
economically done in the framework of external background sources. Notice also
that at order $E^4$ the chiral anomaly can be unambigously included in the
EFT. At order $E^6$, one has to consider loop diagrams with insertions from
${\cal L}_{\rm eff}^{(2)}$ and
${\cal L}_{\rm eff}^{(4)}$ as well as contact terms from
${\cal L}_{\rm eff}^{(6)}$ which introduces new couplings.
Once the low--energy constants are fixed, the aspects of the
dynamics of the standard model related to the chiral symmetry can be worked
out {\it systematically} and {\it unambigously}. Clearly, the EFT can only be
applied below a typical scale $\Lambda \simeq M_\rho$ and higher loop
calculations become more and more cumbersome (but can't always be avoided as
will be discussed below). This is the basic framework of CHPT in a nutshell.
For more details, I refer to refs.[2,3], my review [4] and the extensive list
of references given therein. It is worth pointing out that Leutwyler has
recently given a more sound foundation of the effective Lagrangian approach
by relating it directly to the pertinent  Ward--Identities [5].
\bigskip
\noindent{\bf 2. MESON--MESON SCATTERING AND THE MODE OF QUARK CONDENSATION}
\bigskip
Pion--pion and pion--kaon scattering are the purest reactions between the
pseudoscalar Goldstone bosons. The Goldstone theorem mandates that as the
energy goes to zero, the interaction between the pseudoscalars vanishes.
Consequently, $\pi \pi$ and $\pi K$ scattering are the optimal testing grounds
for CHPT.

Let me first consider the chiral expansion of the isospin zero
S--wave in $\pi \pi$ scattering. In the standard formulation of CHPT, Gasser
and Leutwyler have derived a low--energy theorem generalizing Weinberg's result
(6) [6],
$$a_0^0 = 0.20 \pm 0.01 \eqno(7)$$
which is compatible with the data, $a^0_0 = 0.23 \pm 0.08$ [7]. The theoretical
value (7) rests on the assumption that $B$ is large, i.e. of the order of 1 GeV
(from current values of the scalar quark condensates). However, if $B$ happens
to be small, say of the order of $F_\pi$, one has to generalize the CHPT
framework as proposed by Stern et al.[8]. In that case, the quark mass
expansion of the Goldstone bosons takes the form
$$ M^2_\varphi = m_q \, B + m_q^2 \, A + {\cal O}(m_q^3)  \eqno(8)$$
with the second term of comparable size to the first one.
In ref.[9], this framework is discussed in more detail and a novel
representation of the $\pi \pi$ amplitude which is exact including order $E^6$
and allows to represent the {\it whole} $\pi \pi$ scattering amplitude in terms
of the S-- and P--waves and six subtraction constants is given.
The presently available
data are not sufficiently accurate to disentangle these two possibilities. More
light might be shed on this when the $\phi$--factory DA$\Phi$NE
will be in operation
(via precise measurements of $K_{\ell 4}$--decays, see section 5)
 or if the proposed experiment
to measure the lifetime of pionic molecules [10] will be done. It should also
be pointed out that recent lattice QCD results seem to be at variance with the
expansion (8), but this can only be considered as indicative [11].
Also, the experimentally well--fulfilled GMO relation for the pseudoscalar
meson masses arises naturally in the conventional CHPT framework but requires
parameter fine--tuning in case of a small value of $B \sim 100$ MeV.
Novel high
precision experiments at {\it low} {\it energies} are called for. This is an
important question concerning our understanding of the standard model and it
definitively should deserve more attention. For more details, I refer to
sections 4.1 and 4.2 of ref.[4] as well as ref.[9].

In fig.1, I show the phase--shift $\delta^0_0$ from threshold (280 MeV) to
approximately 600 MeV [12]. One notices the rapid rise of the
phase shift, and at 600 MeV it is already as large as 55 degrees and passes
through 90 degrees at about 850 MeV. At energies below 600 MeV, the other
partial waves do not exceed 15 degrees (in magnitude). This behaviour
of $\delta^0_0$ is attributed to the so--called
strong pionic final state interactions which I will discuss in section 3.

\midinsert
\vskip 10.0truecm
{\noindent\narrower \it Fig.~1:\quad
$\pi \pi$ scattering phase shift $\delta_0^0 (s)$.
The dashed line gives the tree result and the dashed--dotted  the
one--loop prediction. Also shown is the Roy equation band.
The data can be traced back from ref.[12].
The double--dashed line corresponds to the one--loop result
based on another definition of the phase--shift which differs at order $E^6$
from the one leading to the dashed--dotted line (and thus gives a measure of
higher order corrections). On the right side of the hatched area, the one--loop
corrections exceed 50 per cent of the tree result.
\smallskip}
\endinsert

\noindent As indicated in fig.1,
beyond 450 MeV the one loop corrections are half as big
as the tree phase. Nevertheless, one can make a rather precise statement about
the phase of the CP--violation parameter $\epsilon'$[12],
$$\Phi(\epsilon') = {\pi \over 2} -(\delta^0_0 - \delta_0^2)\biggr|_{s =
M_{K^0}^2} = (45 \pm 6)^\circ  \eqno(9)$$
This is due to the fact that the corrections to $\delta_0^2$ are of the same
sign as the ones to $\delta^0_0$ and thus cancel. At tree level, $\Phi
(\epsilon') = 37^\circ$. The accuracy of the theoretical
prediction is as good as the
resent empirical one, $\Phi (\epsilon')_{\rm exp} = (43 \pm 8)^\circ$ [7].
Notice that it is much more difficult to get a precise number on $\Phi
(\epsilon')$ from $K \to 2 \pi$ decays because of the variety of isospin
breaking effects one has to account for (this theme is touched upon in
ref.[12]).

I briefly turn to the case of $\pi K$ scattering. Here, the empirical
situation is even worse, which is very unfortunate. In the framework of
conventional CHPT, the threshold behaviour of the low partial waves can be
unambigously predicted [13] since all low--energy constants in SU(3) are fixed.
Furthermore, since the mass of the strange quark is of the order of the QCD
scale--parameter, it is less obvious that the chiral expansion at
next--to--leading order will be sufficiently accurate.
Much improved empirical
information of these threshold parameters might therefore lead to a better
understanding of the three flavor CHPT. Another possibility is that the
threshold of $\pi K$ scattering at 635 MeV is already so high that one has
to connect CHPT constraints with dispersion theory. This concept has been
investigated in detail by Dobado and Pelaez [14] and certainly improves the
prediction in the P--wave drastically. Another way of extending the EFT through
the implicit inclusion of resonance degrees of freedom is discussed in
ref.[49]. On the experimental side, a measurement of $\pi K$ molecule decays
would certainly help to clarify the situation [11].
\bigskip
\noindent{\bf 3. TWO LOOPS AND BEYOND I: SCALAR FORM FACTOR}
\bigskip
The simplest object to study in detail the strong pionic final state
interactions in the isospin zero S--wave is a three--point function, namely
the so--called scalar form factor (ff) of the pion,
$$ <\pi^a (p') \pi^b (p) | \hat{m} (\bar u u + \bar d d) |0> = \delta^{ab} \,
\Gamma_\pi (s) M_\pi^2    \eqno(10)$$
with $s = (p'+p)^2$. To one loop order, the scalar ff
$\Gamma_{\pi,2} (s)$ has been given in
ref.[3]. As shown in fig.2, closely about the two--pion cut, the real as well
as the imaginary part of the one loop representation are at variance with the
empirical information obtained from a dispersion--theoretical analysis [15].
However, unitarity allows one to write down a two--loop representation [16],
$$\Gamma_\pi (s) = d_0 + d_1 \, s + d_2 \, s^2 + {s^3 \over \pi}
\int_{4M_\pi^2}^\infty {ds' \over s'^3} {\sigma (s') \over s' -s }
\biggl\lbrace
T^0_{0,2} ( 1 + {\rm Re}\Gamma_{\pi,2} ) + T^0_{0,4} \biggr\rbrace \eqno(11) $$
where $T^0_{0,2}$ and $T^0_{0,4}$ are the tree and one loop representations of
the $\pi \pi$ S--wave, isospin zero scattering matrix. Notice that the
imaginary part of $\Gamma_\pi (s)$ to two loops is entirely given in terms of
known one loop amplitudes. The three subtraction constants appearing
in (11) can be fixed from the empirical knowledge of the normalization, the
slope and the curvature of the scalar ff at the origin. In the chiral
expansion, these numbers are combinations of two low--energy constants from
${\cal L}_{\rm eff}^{(4)}$ and two from
${\cal L}_{\rm eff}^{(6)}$.

\midinsert
\vskip 11.0truecm
{\noindent\narrower \it Fig.~2:\quad
Scalar form factor of the pion. The curves labelled
'1', '2', 'O' and 'B' correspond to the chiral prediction to one--loop,
to two--loops, the modified Omn\`es representation and the result of the
dispersive analysis, respectively. The real part is shown in (a) and
the imaginary part in (b).
\smallskip}
\endinsert

The turnover of the scalar ff at around 550 MeV can
be understood if one rewrites (11) in an exponential form,
$$ {\rm Re} \Gamma_\pi (s) = P(s) \exp [{\rm Re}\Delta_0 (s)] \cos\delta^0_0
+ {\cal O}(E^6)   \eqno(12)$$
with Im $\Delta_0 (s) = \delta^0_0 + {\cal O}(E^6)$ fulfilling the final--state
theorem at next--to--leading order. Although this representation is not unique,
it allows to understand the vanishing of Re$\Gamma_\pi (s)$ at 680 MeV since
the phase (in the loop approximation) passes through $90^\circ$ at this energy
thus forcing the turnover. Expanding $\cos \delta^0_0 = 1 - (\delta^0_0)^2 +
\ldots = 1 + {\cal O}(s^2 / F_\pi^4)$ it becomes clear why this behaviour can
only show up at two loop order (and higher). One can do even better and sum up
all leading and next--to--leading logarithms by means of an Omn{\`e}s
representation [16]. This leads to a further improvement in Re$\Gamma_\pi (s)$
and allows to understand that the very accurate two loop result for
Im$\Gamma_\pi (s)$ is not spoiled by higher orders, these can be estimated from
the improved chiral expansion of the scalar ff and are found to be small below
550 MeV. The physics behind all this is that the two--loop corrections lead to
the two--pion cut with proper strength which dominates the scalar ff below 600
MeV. To go further one would have to include inelasticities (which start at
order $E^8$), in particular the strong coupling to the $\bar K K$ channel. It
is also worth pointing out that the scalar ff can only be represented by a
polynomial below $s = 4 M_\pi^2$. Notice that in this energy range the
normalized scalar ff varies from 1 to 1.4, signaling a large scalar radius of
the pion. For comparison, the vector ff  changes from 1 to 1.15 for $0 \le s
\le 4M_\pi^2$. In this way, unitarity allows to extend the  range of CHPT,
however, one has to be able to fix the pertinent subtraction constants (which
is the equivalent to determining the corresponding low--energy constants).
\bigskip
\noindent{\bf 4. TWO LOOPS AND BEYOND II: TWO--PHOTON FUSION}
\bigskip
Another reaction which has attracted much attention recently is $\gamma \gamma
\to \pi^0 \pi^0$ in the threshold region. It belongs to the rare class of
processes which are vanishing at tree level (since the photon can only couple
to charged pions, one needs at least one loop) and do not involve any of the
low--energy couplings from ${\cal L}_{\rm eff}^{(4)}$ at one loop order.
 Some years ago, Bijnens and
Cornet [17] and Donoghue, Holstein and Lin [18] calculated the one--loop cross
section and found that it is at variance with the Crystal ball data [19] even
close to threshold (see fig.3). Denoting by $s$ the cms energy squared, the
amplitude can be written in terms of a single invariant function (at order
$E^4$)
$$\eqalign{
{\cal A}(\gamma \gamma \to \pi^0 \pi^0) &= A(s,t,u) \biggl[ -{s \over 2}
\epsilon_1 \cdot \epsilon_2 + \epsilon_1 \cdot k_2 \epsilon_2 \cdot k_1 \biggr]
\cr  A(s,t,u) &= A^\pi (s,t,u) + A^K (s,t,u)  \cr
A^\pi (s,t,u) &= i e^2 {1 \over 4 \pi^2 F_\pi^2} \biggl[ 1 - {M_\pi^2 \over 2}
\biggr] \biggl[ 1 + {M_\pi^2 \over s} \ln^2 Q_\pi \biggr] \cr
A^K (s,t,u) &= i e^2 {1 \over 16 \pi^2 F_\pi^2}
 \biggl[ 1 + {M_K^2 \over s} \ln^2 Q_K \biggr] \cr
Q_i &= {\sqrt{s_i -4} + \sqrt{s_i} \over \sqrt{s_i -4} - \sqrt{s_i}} \, , \quad
s_i = {s \over M_i^2} \, (i= \pi,K) \cr}
\eqno(13)$$
with the contribution to the pion loops being completely dominant. The apparent
discrepancy between the one--loop prediction and the data (cf. Fig.3) even very
close to threshold was long considered a severe problem for CHPT. Notice that
for charged pion production, this problem does not occur since there is a
dominant ${\cal O}(E^2)$ contribution which already is close to the data.
Furthermore, in the threshold region the total cross section for $\gamma \gamma
\to \pi^+ \pi^-$ is approximately two order of magnitude larger than for the
double neutral pion production, i.e. one is after a small effect.
\midinsert
\vskip 10.0truecm
{\noindent\narrower \it Fig.~3:\quad
Cross section for $\gamma \gamma \to \pi^0 \pi^0$. The chiral one and two loop
predictions are given by the dotted and the solid line, in order. The hatched
area is a dispersion--theoretical fit. The Crystal ball data are also shown.
{}From [22].\smallskip}
\vskip -0.5truecm
\endinsert

In fact, $\gamma \gamma \to \pi^0 \pi^0$ is another case
where one has to account for the
strong pionic final state interactions. At 400 MeV, one has
$$\biggl({\sigma^{\exp} \over \sigma^{\rm 1-loop}}\biggr)^{1/2}
= 1.3 \eqno(14)$$
which is a typical  correction in this channel (see discussion above
on $a_0^0$ and the scalar ff).
In fact, dispersion theoretical
calculations supplemented with current algebra constraints by Pennington [20]
tend to give the trend of the data (see the shaded area in fig.3). An improved
combination of chiral machinery and dispersion theory has been given by
Donoghue and Holstein [21]. Even better, Bellucci, Gasser and Sainio [22] have
performed a full two loop calculation. It involves some massive algebra and
three new low--energy constants have been estimated from resonance exchange
(the main contribution comes form the $\omega$). These couplings play, however,
no role below 400 MeV. The solid line in fig.3 shows the two--loop result for
the central values of the coupling constants. One finds a good agreement
with the data up to $E_{\pi \pi} = 700$ MeV. This resolves the long--standing
discrepancy between the chiral prediction and the data in the threshold region.
For a more detailed discussion of these topics and the related neutral pion
polarizabilities, I refer to ref.[22].

Another topic I briefly want to mention in connection with large unitarity
corrections is the radiative kaon decay $K_L \to
\pi^0 \gamma \gamma$ which has no tree--level contribution and is given by a
finite one--loop calculation at order $E^4$ [23]. The predicted two--photon
invariant mass spectrum turned out to be in amazing agreement with the later
measurements [24].
However, the branching ratio which is also predicted was found
 about a  factor three too
small. Again, unitarity corrections work in the right direction. In recent
work by D'Ambrosio and collaborators [25] and later by
Cohen, Ecker and Pich [26] as well as Kambor and Holstein [27] it is shown that
unitarity corrections (eventually supplemented by a sizeable $E^6$ vector meson
exchange contribution)
can indeed close the gap between the empirical branching
ratio and the CHPT prediction though not completely. These calculations are,
however, not taking into account all effects beyond $E^4$ but they underline
the importance of making use of dispersion theory in connection with CHPT.
\vfill \eject
\bigskip
\noindent{\bf 5. TWO LOOPS AND BEYOND III: $K_{\ell 4}$--DECAYS}
\bigskip
As already mentioned, $K_{\ell 4}$--decays give information concerning the
$\pi \pi$ phase shifts close to threshold (for a good but somewhat old review,
see ref.[28]). Because of the $\Delta I = 1/2$ rule the two pions in the final
state can only have total isospin zero or one. In principle, the energy of the
two pions $\sqrt{s_\pi}$
lies between $2M_\pi$ and $M_K - m_l$, with $m_l$ the mass of the
corresponding lepton (which in the case of the electron can mostly be
neglected). Due to phase space, however, only the first 100 MeV above two--pion
threshold are really available and one is thus sensitive to the phase
difference $\delta_0^0 (s_\pi) - \delta_1^1 (s_\pi)$ and $\delta_1^1 (s_\pi)$
stays below 2$^\circ$ for $\sqrt{s_\pi} < 380$ MeV, i.e. one essentially
measures the isospin zero, S--wave. In the more refined treatment discussed
below, one takes into account all partial waves allowed. The largest data
sample presently available are the 30000 $K_{e4}$ decays measured and analyzed
by the CERN-Geneva group [29]. I refer to that reference for a detailed account
of the measured distributions and conclusions drawn at that time.

To be specific, consider the decay $K^+ \to \pi^+ \pi^- \ell^+ \nu_\ell$. The
transition matrix--element factors into a leptonic times a hadronic current
$$ T = {G_F \over \sqrt{2}} V^*_{us} \bar{u}(p_\ell ) \gamma_\mu (1 - \gamma_5
) \nu (p_\nu) <\pi^+ (p_1) \pi^- (p_2) | I_\mu^{4 -i5} (0)|K^+ (p)> \, \quad
I = V,A  \eqno(15)$$
with $G_F$ the Fermi constant, $V_{us}$ the pertinent entry in the CKM matrix
and the leptonic current is completely known. In contrast, the hadronic ME
is parametrized in terms of four form factors, three related to the
axial--vector current $A_\mu$ (denoted $F,G$ and $R$) and one related to the
vector current $V_\mu$ (denoted $H$)
$$\eqalign{
V_\mu &= - {H \over M_K^3} \epsilon_{\mu \nu \rho \sigma} (p_\ell + p_\nu)^\nu
(p_1 + p_2)^\rho + (p_1 - p_2)^\sigma \cr
A_\mu &= - {i \over M_K^3} \biggl[ (p_1 + p_2)_\mu F + (p_1 - p_2)_\mu G +
(p_\ell + p_\nu)_\mu R \biggr] \cr}  \eqno(16)$$
Clearly, these form factors contain the hadronic physics. The ff $H$ is
obvioulsy related to the chiral anomaly and will be discussed later (since
the ME of the vector current is proportional to the totally antisymmetric
tensor in four dimensions). The ff $R$ is only relevant for heavy leptons,
say for the muon. I will not discuss it in what follows.

The chiral expansion for the ffs $F,G$ and $H$ to leading order $E^2$ was first
given by Weinberg [30] and reads
$$F = G = {M_K \over \sqrt{2} F_\pi} = 3.74 \, \quad H = 0   \eqno(17)$$
i.e. at that order one sees no momentum dependence (cf. the discussion of the
scalar ff of the pion in section 3). The one--loop representation
for $F$ takes the form [31,32]
$$F(s_\pi,t,u) = {M_K \over \sqrt{2} F_\pi} \biggl[ 1 + {1 \over F_\pi^2} (U_F
+ P_F + C_F ) \biggr] + {\cal O}(E^6)  \eqno(18)$$
with $t = (p-p_1)^2 , u = (p-p_2)^2$ and the expansion for $G$ looks similar.
At next--to--leading order, one has unitarity corrections ($U_F$) from
the one loop graphs, in this particular case they are
proportional to the tree level
prediction for $\delta_0^0 (s_\pi)$. The low--energy constants $L_i$ are
subsumed in the polynomial piece $P_F$ and $C_F$ contains chiral logs. As it
turns out, these ffs are only sensitive to $L_{1,2,3}$ and thus the other
$L_i's$ contributing are taken from previous determinations. The $E^4$
prediction for $H$ will be discussed later.

Now we can ask the question whether
this one loop result for the ffs $F$ and $G$ will be sufficiently accurate to
pin down the low--energy constants $L_{1,2,3}$ and therefore the $\pi \pi$
phases? For that, we compare with the data of ref.[29] at threshold,
$$F^{\rm thr} = 3.74 [ 1 + m_q ( \alpha + \beta L_1 + \gamma L_2 + \delta L_3)
+ \ldots ] \krig{=}   5.59 \pm 0.14      \eqno(19)$$
where the ellipsis stands for one and higher loop contributions. Clearly, the
empirical number is  a factor 1.5 larger than the tree prediction. The $L_i$
can only be determined precisely if one can estimate the higher order
corrections. This has been done by Bijnens, Colangelo and Gasser [33] who
write down a dispersive representation for $F$ (and also $G$), $F = f_s \,
\exp(i \delta^0_0 )$ in the spirit of section 3. The two--pion cut is taken out
by a modified Omn{\`e}s function and the remaining polynomial piece is smooth,
$\tilde{f}_s = a + b m_q$ and terms of order $m_q^2$ have been neglected. The
details of this procedure are spelled out in ref.[33]. While the resulting
numbers for the $L_{1,2,3}$ are sensitive to the inclusion of higher orders,
one finds a beautiful consistency between the $\pi \pi$ threshold parameters
derived first only from the $K_{e4}$ data of ref.[29] and second by adding the
existing threshold $\pi \pi$ data from other reactions. This is shown in table
1 together with the empirical numbers from Petersen [34]. It should also be
stressed that the one--loop plus unitarization calculation leads to
a much improved description of the $\pi \pi$ D--wave scattering lengths. These
were originally used to pin down the values for $L_1$ and $L_2$ [3,6] and that
procedure was often criticized since the empirical values have large
uncertainties. In ref.[33], many other chiral predictions are given and I
refer the reader for all the details to that paper.
$$\hbox{\vbox{\offinterlineskip
\def\strut{\hbox{\vrule height 12pt depth 12pt width 0pt}}
\hrule
\halign{
\strut\vrule# \tabskip 0.1in &
\hfil#\hfil &
\vrule# &
\hfil#\hfil &
\hfil#\hfil &
\hfil#\hfil &
\vrule# \tabskip 0.0in
\cr
&   &&  $K_{e4}$ data & $K_{e4} \, + \, \pi\pi$ data & Exp. & \cr
\noalign{\hrule}
& $ a_0^0 $        && 0.20 & 0.20 & $0.26 \pm 0.05$ & \cr
& $ -10 \, a_0^0 $ && 0.41 & 0.41 & $0.28 \pm 0.12$ & \cr
& $ 10 \, a_1^1 $  && 0.37 & 0.37 & $0.38 \pm 0.02$ & \cr
& $ 100 \, a_2^0 $ && 0.18 & 0.18 & $0.17 \pm 0.03$ & \cr
& $ 100 \, a_2^2 $ && 0.21 & 0.20 & $0.13 \pm 0.30$ & \cr
\noalign{\hrule}}}}$$
\smallskip
{\noindent\narrower \it Table 1:\quad $\pi \pi$ scattering lengths in
appropriate units of inverse pion masses. The numbers are taken from ref.[33]
and represent the calculation including higher loop effects via unitarization.
\smallskip}
\bigskip
\noindent{\bf 6. ANOMALIES: GENERAL REMARKS}
\bigskip
As already mentioned, there are processes which are proportional to the totally
antisymmetric tensor in four dimensions. These are related to anomalies, in our
case the so--called chiral anomaly. In this section, I will give a short
discussion
about the meaning of anomalies in QFTs. For more details, I refer to the
monograph by Treiman, Jackiw, Zumino and Witten [35] (and refs. therein).

First, I have to define what an anomaly is. One speaks of an {\it anomaly} if a
{\it classical} {\it Lagrangian} {\it symmetry} is {\it broken} {\it upon} {\it
quantization}. Although anomalies are related to short distance phenomena,
they show up most clearly at long wave lenghts (as I will show in what
follows). Furthermore, in QFTs such effects are quite normal, remember that
anomaly cancelation plays a central role in the quantization of field theories
like the standard model. To get an idea, let me briefly give a field theoretic
view of anomalies in the path integral formalism following the work of Fujikawa
[36]. Consider a Lagrangian  ${\cal L}(\Psi, \bar{\Psi}, \ldots )$
which is invariant under transformations like
$$\Psi \to \Psi' = \exp[iS] \, \, \Psi \, , \ldots    \eqno(20)$$
with $S = S_a T_a$ and $T_a$ the generators of the correspnding algebra, i.e.
${\cal L}(\Psi', \bar{\Psi}', \ldots ) = {\cal L}(\Psi, \bar{\Psi}, \ldots )$.
At the quantum level, we consider the generating functional
$$ {\cal Z} = \int [d\Psi] [d\bar{\Psi}] [\ldots]
\exp \biggl\lbrace i \int d^4 x
{\cal L}(\Psi, \bar{\Psi}, \ldots )  \biggr\rbrace \, \, .    \eqno(21)$$
Clearly, under the symmetry transformation related to $S$ the measure might
change,
$$ [d\Psi] [d\bar{\Psi}]
\to [d\Psi'] [d\bar{\Psi}']  |{\cal J}|   \eqno(22)$$
so if the Jacobian is not equal one, $|{\cal J}| \neq 1$,
we encounter an anomaly, i.e. the
classical Lagrangian symmetry is broken. As an example, consider massless QED
where $\Psi$ denotes an isodoublet (say of u and d quarks) and $A_\mu$ the
photon field (U(1) gauge field),
$${\cal L}(\Psi, \bar{\Psi}, A_\mu) = \bar{\Psi} ( i \barre{\partial} - Q
A_\mu) \Psi - {1 \over 4} F_{\mu \nu} F^{\mu \nu}      \eqno(23)$$
where $F_{\mu \nu} = \partial_\mu A_\nu - \partial_\nu A_\mu$
is the photon field strength tensor and $Q$ is the (quark) charge matrix.
Under axial transformations $\Psi \to \exp[i \kappa \gamma_5]
\, \Psi$ the Lagrangian is obviously invariant and the corresponding Noether
current $J^{\mu 5} = \bar{u} \gamma_\mu \gamma_5 u -  \bar{d} \gamma_\mu
 \gamma_5 d$ is conserved, $\partial_\mu J^{\mu 5} = 0$. Upon quantization, the
measure picks up a nontrivial Jacobian,
$ [d\Psi'] [d\bar{\Psi}'] = [d\Psi] [d\bar{\Psi}] \exp[-2i {\rm Tr} ( \kappa
\gamma_5)]$ which leads to a non--vanishing derivative of the axial current,
$$\partial_\mu J^{\mu 5} = {\alpha \over 12 \pi} N_c \epsilon_{\mu \nu \lambda
\sigma} F^{\mu \nu} F^{\lambda \sigma}               \eqno(24)$$
with $\alpha = 1/137$ the fine structure constant and $N_c$ denotes the number
of colors. This is, indeed, the original Adler--Bell--Jackiw [37] anomaly which
leads to a finite lifetime for the $\pi^0$ decay into two photons,
$$\Gamma (\pi^0 \to 2 \gamma) = {\alpha^2 N_c^2 M_\pi^3 \over 576 \pi^3
F_\pi^2} = (7.6 \, {\rm eV}) \, {N_c^2 \over 9}   \eqno(25)$$
which compared with the empirical value of $(7.7 \pm 0.6)$ eV is one of the
strongest arguments that the number of colors is indeed three, $N_c = 3$.
\bigskip
\noindent{\bf 7. THE CHIRAL ANOMALY A LA WESS--ZUMINO--WITTEN}
\bigskip
The chiral anomaly was first discussed by Wess and Zumino [38] in the context
of anomalous Ward identities and later given a beautiful geometrical
interpretation by Witten [39]. I will essentially only give a pedagogical
treatment of the topic following the review [40]. To be specific, let us
consider the first term in the energy expansion of eq.(4),
$${\cal L}^{(2)} = {F_\pi^2 \over 4} \Tr ( \partial_\mu U \partial^\mu
U^\dagger )                  \eqno(26)$$
where $U(x)$ is an element of SU(3) and subsumes the Goldstones (there is no
chiral anomaly for the two flavor case). As is QCD, ${\cal L}^{(2)}$ is
invariant under parity, $ P U(\vec{x} , t)P^{-1} = U^\dagger (-\vec{x} , t)$.
However, besides that, ${\cal L}^{(2)}$ has two extra symmetries, i.e. it is
invariant under
$$U(\vec{x} , t) \to U(-\vec{x} , t) \quad {\rm and} \quad
U(\vec{x} , t) \to U^\dagger(\vec{x} , t) \quad .       \eqno(27)$$
It is easily understood that this means that intrinsic parity or the number of
Goldstone bosons modulo two is conserved. Intrinsic parity is defined as
follows. For a true (pseudo) tensor of rank k, intrinsic parity $P_I$ is plus
(minus) one. So scalars, polar vectors, $\ldots$ have $P_I = +1$ whereas
pseudoscalars, axial vectors, $\ldots$ have $P_I =-1$. Furthermore, intrinsic
parity is a multiplicative quantum number. Typical processes conserving $P_I$
(and thus the number of Goldstones modulo two) are $\pi \pi \to \pi \pi$,
$\gamma \pi \to \pi$, $\gamma \gamma \to \pi \pi$ or $\eta \to 3 \pi$.
Intrinsic parity is violated for $\pi^o \to 2 \gamma$, $K^+ K^- \to \pi^+ \pi^-
\pi^0$, $\gamma \to \pi^+ \pi^- \pi^0$ or the ME $<\pi \pi | V_\mu| K>$
encountered in section 5. Similar observations can be made for the terms of
order $E^4$ (and higher) in eq.(4). So the moral is that while the effective
Lagrangian does conserve $P_I$, QCD does not. To break the redundant
symmetries on the level of the equation of motion for $L_\mu = U^\dagger
\partial_\mu U$ derived from ${\cal L}^{(2)}$,
one can easily write down an extra term [39],
$${i \over 2} F_\pi \partial^\mu L_\mu + \lambda \epsilon^{\mu \nu \alpha
\beta} L_\mu L_\nu L_\alpha L_\beta + \ldots = 0           \eqno(28)$$
with the constant $\lambda$ to be fixed later. As pointed out by Witten, this
can not be written in terms of a four--dimensional Lagrangian but rather
as an integral over a 5--dimensional sphere which bounds space--time (I write
down only the term for the interactions between the Goldstone bosons),
$$ \Gamma_{\rm WZW} = -{i n \over 240 \pi^2} \int_{S^5} d^5x \, \,
 \epsilon^{\mu \nu \alpha \beta \gamma} \Tr [ L_\mu L_\nu L_\alpha L_\beta
L_\gamma ] \quad .                    \eqno(29)$$
By topology, $n$ has to be an integer number. It can be fixed when one gauges
$ \Gamma_{\rm WZW}$ (correctly done first in refs.[41]) and compares with the
result for $\pi^0 \to 2 \gamma$, eq.(24). This leads to the identification
$$ n = N_c      \eqno(30)$$
and therefore the effective meson Lagrangian still knows about the number of
colors of QCD, an amazing result. From the point of chiral counting, the
Wess--Zumino--Witten term is of order $E^4$. At this leading order, it is
uniquely fixed, i.e. does not introduce any novel low--energy constant. Let me
finish this section by one curious experimental result. From the gauged WZW
action, one can immediately derive the amplitude for $\gamma \to 3 \pi$, $
A(\gamma \to 3 \pi) = e N_c
/ 12 \pi^2 F_\pi = 9.7$ GeV$^{-3}$. In ref.[42], an empirical determination of
this quantity was presented. The measurement was based on the production of
pions in the virtual field of a nuclues, $\pi A \to \pi \pi A$ via one photon
exchange. Interpolating to the low energy limit, one arrived at $12.9 \pm 0.9
\pm 0.5$ GeV$^{-3}$ [42] which is at variance with the theoretical prediction.
However, a remeasurement as well as more thorough theoretical calculations are
called for before one can draw a final conclusion. In fact, if one considers
the process $\gamma N \to \pi \pi N$ (here, $N$ denotes the nucleon) [43],
 there are many other competing
diagrams and it is not yet clear how cleanly one could separate out the
anomalous $\gamma 3 \pi$ vertex.
\bigskip
\noindent{\bf 8. SIGNALS OF THE CHIRAL ANOMALY}
\bigskip
In this section, I will briefly talk about a few intrinsic parity violating
reactions related to decays of pions, kaons and etas. An older review is
ref.[44] and a fresh look in view of CHPT has recently been given by Bijnens
[45] (which contains much more details than given here).

A first example is the decay mode $\eta \to \pi^+ \pi^- \gamma$. From the
WZW action, one predicts $\Gamma (\eta \to \pi^+ \pi^- \gamma ) = 35$ MeV
to be compared with the PDG value of $58 \pm 6$ eV. However, there is large
vector meson contribution starting at order $E^6$ which  goes in the right
direction. Now let me return to the $K_{e4}$--decays discussed previously.
As already noted, the contribution of the vector current to the hadronic ME is
of anomalous nature and thus only starts to contribute at order $E^4$ and is
entirely given in terms of known parameters [31,32]
$$ H = - {\sqrt{2} M_K^3 \over 8 \pi^2 F_\pi^3} + {\cal O}(E^6) = -2.66
\eqno(31)$$
which compares well with the empirical number [29],
$$ H_{\rm exp} = -2.68 \pm 0.68      \eqno(32) $$
if one uses the pion decay constant. It would also be legitimate to use $F_K
=1.22 \, F_\pi$ here, thus reducing the theoretical prediction by a factor 1.8.
The order $E^6$ corrections have been calculated and found to be small if one
estimates the appearing low--energy constants using vector mesons only [46].
 Most
amazing, however, is the fact that the empirical result checks indeed the sign
of the chiral anomaly. Another wide field to study the chiral anomaly are
radiative pion and kaon decays such as $\pi \to e \nu_e \gamma$, $K \to l
\nu_l \gamma$ or $K \to \pi l \nu_l \gamma$. For a detailed discussion of
these, I refer the reader to the updated version of the DA$\Phi$NE handbook
[47]. Finally, I wish to mention  non--leptonic radiative $K$--decays.
Examples are $K_L \to \pi^+ \pi^- \gamma$, $K^+ \to \pi^+ \pi^0 \gamma$,
$K \to \pi \pi \gamma \gamma$ or $K \to 3 \pi \gamma (\gamma )$. These have
been studied extensively by Ecker, Neufeld and Pich [48] (see also the
references given therein). Apart from the reducible amplitudes, which can
directly be derived from the WZW functional (and are thus unambigous), there
are also so--called direct contributions which induce some theoretical
uncertainties. In certain channels, one furthermore has to account for the
$E^6$ contributions, which come form $\eta-\eta'$ mixing and vector meson
exchange. This is a complementary field of testing the chiral anomly, which is
and will be exploited in more detail in the future. I finish this section
with the remark that
at present the "standard" anomalous strong process, namely $K^+ K^- \to
\pi^+ \pi^- \pi^0$ has not yet been observed.
\bigskip
\noindent{\bf 9. WHY?}
\bigskip
In this lecture I could only give a glimpse of the many facets of chiral
perturbation theory. Instead of repeating what was already said, let me
briefly remind you of why all these calculations are done. Clearly, the
effective chiral Lagrangian approach gives us some insight about some
{\it fundamental} {\it parameters} of {\it QCD}. In the
introduction I mentioned already the
ratios of the light quark masses and in section 2 I pointed out that refined
measurements of e.g. the $\pi \pi$ threshold parameters would put stringent
test on our understanding of the mode of quark condensation,
in particular how large
the value of the order parameter $B$ actually is.
Furthermore, the chiral anomaly is a
direct consequence of the fact that the standard model is a chiral QFT. At
present, not too many experimental tests of this important ingredient of modern
particle physics exist. CHPT is the effective field theory of the Standard
Model at low energies and has to be subjected to as many empirical tests as
possible. In general, calculations to order $E^4$ are already  accurate,
however, as discussed here, there exist circumstances when one has to work
harder. These are essentially related to strong pionic final state interactions
and can be treated in a combination of dispersion theory with CHPT constraints.
We are looking forward to the operation of DA$\Phi$NE and it might also be
worthwhile to analyze the many K--decays which are on tape from other
experiments but are only considered as backgrounds.
\bigskip \bigskip
\noindent I would like to thank the organizers for their invitation
and the efficient organization.
\vfill \eject
\noindent{\bf REFERENCES}
\medskip
\item{1.}S. Weinberg, {\it Phys. Rev. Lett.\/} {\bf 17} (1966) 616.
\smallskip
\item{2.}S. Weinberg, {\it Physica} {\bf 96A} (1979) 327.
\smallskip
\item{3.}J. Gasser and H. Leutwyler, {\it Ann. Phys. {\rm (N.Y.)}\/}
 {\bf 158} (1984) 142;  {\it Nucl. Phys.\/} {\bf B250} (1985) 465.
\smallskip
\item{4.}Ulf-G. Mei{\ss}ner, {\it Rep. Prog. Phys.\/} {\bf 56} (1993) 903.
\smallskip
\item{5.}H. Leutwyler, {\it Ann. Phys.} (N.Y.) (1994), in print. \smallskip
\item{6.}J. Gasser and H. Leutwyler,
{\it Phys. Lett.\/} {\bf 125B} (1983) 325.
\smallskip
\item{7.}W. Ochs, Max--Planck--Institute preprint MPI--Ph/Ph 91--35, 1991.
\smallskip
\item{8.}N.  H. Fuchs, H. Szadijan and J. Stern, {\it Phys. Lett.\/} {\bf
B269} (1991) 183. \smallskip
\item{9.}N.  H. Fuchs, H. Szadijan and J. Stern, {\it Phys. Rev.\/} {\bf
D47} (1993) 3814. \smallskip
\item{10.}L. L. Nemenov, {\it Yad. Fiz.\/} {\bf 41} (1985) 980;

J. Uretsky and J. Palfrey, {\it Phys. Rev.\/} {\bf 121} (1961) 1798;

G. Czapek et al., letter of intent CERN/SPSLC 92--44, 1992.
\smallskip
\item{11.}R. Altmeyer et al., J\"ulich preprint HLRZ 92--17, 1992.
\smallskip
\item{12.}J. Gasser and Ulf-G. Mei{\ss}ner, {\it Phys. Lett.\/} {\bf B258}
(1991) 219.
\smallskip
\item{13.}V. Bernard, N. Kaiser and Ulf-G. Mei{\ss}ner,
{\it Nucl. Phys.\/} {\bf B357} (1991) 129;
{\it Phys. Rev.\/} {\bf D43} (1991) R3557.
\smallskip
\item{14.}A. Dobado and J. R. Pelaez,
{\it Phys. Rev.\/} {\bf D47} (1993) 4883. \smallskip
\item{15.}J. F. Donoghue, J. Gasser and H. Leutwyler, {\it Nucl. Phys.\/}
{\bf B343} (1990) 341.
\smallskip
\item{16.}J. Gasser and Ulf-G. Mei{\ss}ner,
{\it Nucl.Phys.\/} {\bf B357} (1991) 90.
\smallskip
\item{17.}J. Bijnens and F. Cornet, {\it Nucl. Phys.\/} {\bf B296} (1988) 557.
\smallskip
\item{18.}J. F. Donoghue, B. R. Holstein and Y. C. R. Lin,  {\it Phys. Rev.
\/} {\bf D37} (1988) 2423.
\smallskip
\item{19.}H. Marsiske et al.,  {\it Phys. Rev.\/} {\bf D41} (1990) 3324.
\smallskip
\item{20.}M.R. Pennington, in The DA$\Phi$NE Physics Handbook, eds. L. Maiani,
G. Pancheri and N. Paver, Frascati, 1992. \smallskip
\item{21.}J. F. Donoghue and B. R. Holstein,
 {\it Phys. Rev.\/} {\bf D48} (1993) 37. \smallskip
\item{22.}S. Belluci, J. Gasser and M.E. Sainio, Bern University preprint
BUTP-93/18, 1993. \smallskip
\item{23.}G. Ecker, A. Pich and E. de Rafael, {\it Phys. Lett.\/} {\bf
B189} (1987) 363; {\it Nucl. Phys.\/} {\bf B291} (1987) 692;
{\it Nucl. Phys.\/} {\bf B303} (1988) 665;

L. Cappiello and G. D'Ambrosio, {\it Nuovo Cim.\/} {\bf 99} (1988) 155.
\smallskip
\item{24.}G.D. Barr et al.,  {\it Phys. Lett.\/} {\bf B242} (1990) 523;
{\bf B284} (1992) 440;

V. Papadimitriou et al., {\it Phys. Rev.\/} {\bf D44} (1991) 573.
\smallskip
\item{25.}L. Cappiello, G. D'Ambrosio and M. Miragliulio,  {\it Phys. Lett.\/}
{\bf B298} (1993) 423.  \smallskip
\item{26.}A. Cohen, G. Ecker and A. Pich,  {\it Phys. Lett.\/}
{\bf B304} (1993) 347.  \smallskip
\item{27.}J. Kambor and B.R. Holstein, Amherst preprint UMHEP--397, 1993.
\smallskip
\item{28.}L.-M. Chounet, J.-M. Gaillard and M.K. Gaillard,
{\it Phys. Rep.} {\bf C4} (1972) 199.
\smallskip
\item{29.}L. Rosselet et al., {\it Phys. Rev.\/} {\bf D15} (1977) 574.
\smallskip
\item{30.}S. Weinberg, {\it Phys. Rev. Lett.\/} {\bf 17} (1966) 366; {\bf 18}
(167) 1178(E). \smallskip
\item{31.}J. Bijnens, {\it Nucl. Phys.\/} {\bf B337} (1990) 635.
\smallskip
\item{32.}C. Riggenbach, J. Gasser, J.F. Donoghue and B.R. Holstein,
{\it Phys. Rev.} {\bf D43} (1991) 127.   \smallskip
\item{33.}J. Bijnens, G. Colangelo and J. Gasser, BUTP--94/4, in print
 \smallskip
\item{34.}J.L. Petersen, CERN yellow report 77--04, 1977.  \smallskip
\item{35.}S.B. Treiman, R. Jackiw, B. Zumino and E. Witten, "Current Algebras
and Anomlies", Princeton University Press, Princeton, 1985. \smallskip
\item{36.}K. Fujikawa, {\it Phys. Rev.\/} {\bf D21} (1980) 2848.
\smallskip
\item{37.}S.L. Adler,
{\it Phys. Rev.\/} {\bf 177} (1969) 2426;

J.S. Bell and R. Jackiw, {\it Nuovo Cim.} {\bf 60A} (1969) 47. \smallskip
\item{38.}J. Wess and B. Zumino,
{\it Phys. Lett.\/} {\bf B37} (1971) 95. \smallskip
\item{39.}E. Witten, {\it Nucl. Phys.\/} {\bf B223} (1983) 422.
\smallskip
\item{40.}Ulf-G. Mei{\ss}ner and I. Zahed, {\it Adv. Nucl. Phys.} {\bf 17}
(1987) 143. \smallskip
\item{41.}K. Chou et al., {\it Phys. Lett.} {\bf B134} (1984) 67;

H. Kawai and S. Tye, {\it Phys. Lett.} {\bf B140} (1984) 403;

O. Kaymakcalan et al., {\it Phys. Rev.\/} {\bf D30} (1984) 594;

J. Manes, {\it Nucl. Phys.} {\bf B250} (1985) 369.  \smallskip
\item{42.}Yu M. Antipov et al.,
{\it Phys. Rev.\/} {\bf D36} (1987) 21.   \smallskip
\item{43.}V. Bernard, N. Kaiser, Ulf-G. Mei{\ss}ner and A. Schmidt,
{\it Nucl. Phys.\/} {\bf A590} (1994) in print.  \smallskip
\item{44.}Ulf-G. Mei{\ss}ner, {\it Phys. Rep.} {\bf 161} (1988) 213.
\smallskip
\item{45.}J. Bijnens, {\it Int. J. Mod. Phys.} {\bf A8} (1993) 3045.
\smallskip
\item{46.}Ll. Ametller et al., {\it Phys. Lett.\/}
 {\bf B303} (1993) 140. \smallskip
\item{47.}The DA$\Phi$NE Physics Handbook, second edition, in preparation.
\smallskip
\item{48.}G. Ecker, H. Neufeld and A. Pich,
{\it Nucl. Phys.\/} {\bf B413} (1994) 321. \smallskip
\item{49.}V. Bernard, N. Kaiser and Ulf--G. Mei{\ss}ner, {\it Nucl. Phys.\/}
{\bf B364} (1991) 283. \smallskip
\vfill \eject \end